\documentclass{aastex}
\usepackage{spr-astr-addons}             
\usepackage{url}
\usepackage{lscape, morefloats, graphicx, float, longtable}   
\RequirePackage{color}                   

\begin{document}

\title{Photometric Calibration of the [$\alpha$/Fe] Element:\\
 I. Calibration with $UBV$ Photometry}
\slugcomment{Not to appear in Nonlearned J., 45.}
\shorttitle{I. Calibration with $UBV$ Photometry}
\shortauthors{S. Karaali, E. Yaz G\"ok\c ce, S. Bilir}

\author{S. Karaali \altaffilmark{1}}
\altaffiltext{1}{Istanbul University, Faculty of Science, Department 
of Astronomy and Space Sciences, 34119 University, Istanbul, Turkey\\
\email{karsa@istanbul.edu.tr}}

\author{E. Yaz G\"ok\c ce \altaffilmark{1}} 
\altaffiltext{1}{Istanbul University, Faculty of Science, Department 
of Astronomy and Space Sciences, 34119 University, Istanbul, Turkey\\}

\and
\author{S. Bilir \altaffilmark{1}}
\altaffiltext{1}{Istanbul University, Faculty of Science, Department 
of Astronomy and Space Sciences, 34119 University, Istanbul, Turkey\\}

\begin{abstract}  
We present the calibration of the [$\alpha$/Fe] element in terms of the ultra-violet 
excess for 469 dwarf stars with $0.325<(B-V)_0 \leq 0.775$ mag corresponding the 
spectral type range F0-K2. The star sample is separated into nine sub-samples with 
equal range in $(B-V)_0$ colour, $\Delta(B-V)_0=0.05$ mag, and a third degree 
polynomial is fitted to each dataset. Our calibrations provide [$\alpha$/Fe] 
elements in the range [0.0, 0.4]. We applied the procedure to two sets of field 
stars and two sets of clusters. The mean and the corresponding standard deviation 
of the residuals for 43 field stars taken from the Hypatia catalogue are 
[$\alpha$/Fe]=-0.090 and $\sigma =0.102$ dex, while for the 39 ones taken from the 
same catalogue of stars used in the calibration are [$\alpha$/Fe]=-0.009 and 
$\sigma = 0.079$ dex, respectively. We showed that the differences between the mean 
of the residuals and standard deviations for two sets of clusters ([$\alpha$/Fe]=0.073 
and $\sigma = 0.91$ dex; [$\alpha$/Fe]=-0.012 and $\sigma = 0.053$ dex) originate 
from the $(B-V)_0$ and $(U-B)_0$ colour indices of the clusters which are taken 
from different sources. The differences between the original [$\alpha$/Fe] elements 
and the estimated ones (the residuals) are compatible with the uncertainties in the 
literature. Also, there is a good agreement between the distribution of the synthetic 
alpha elements versus ultra-violet excesses and the ones obtained via our calibrations. 

\end{abstract}

\keywords{(stars:) late-type - dwarfs - abundances  (techniques:) 
photometric (methods:) statistical\\
(Galaxy:) Globular clusters}

\section{Introduction}

Metallicity is one of the means to investigate the formation and evolution of our Galaxy. 
Metallicity of a star can be determined by spectroscopic measures of its surface, and 
it is thought to represent the chemical composition of the gas cloud that collapsed 
to form it. Different metallicity abundances reveal the formation time of that star. 
Hydrogen (H) and helium (He) were produced after the Big-bang, while metals (elements 
heavier than H and He) are products of nuclear reactions interior of the stars. The metals 
are produced from different sources. Alpha ($\alpha$) elements which are formed 
by the fusion of alpha particles ($^{4}$He-nuclei) and iron element (Fe) are produced 
from Type II supernovae on short timescales, i.e. 20 Myr \citep{Wyse88}, while iron-peak 
elements are produced mainly from Type Ia supernovae on much larger timescales, i.e. 
a few Gyr. The increase of Fe element by time reduces the [$\alpha$/Fe] value of 
the interstellar cloud that produces new stars. Hence, one expects a flat distribution 
for metal poor and old stars and a negative gradient for relatively metal-rich stars. 
The relation between the alpha element relative to the iron element, [$\alpha$/Fe], 
in terms of classical iron element, [Fe/H], is a universal clock for investigation 
of the formation and evolution of our Galaxy. 

The interpretation of \citet{Roman55} of the weakness of the metallicity lines in the 
F- and G- type spectra by comparison of the $B-V$ and $U-B$ colours for each star could 
be reduced to a procedure for  metallicity determination of a star, i.e. the metallicity 
of a star can be measured by its ultra-violet excess, $\delta(U-B)$, which is defined 
as the difference between the $U-B$ colours of a star and a standard one with the same 
$B-V$ colour. It has been a custom to use the $U-B \times B-V$ two-colour diagram of 
the Hyades cluster as a standard sequence in ultra-violet excess determinations. However, 
this procedure should be applied with caution due to the guillotin effect which reduces 
the ultra-violet excess of the red stars with the same metallicity of bluer ones. 
\citet{Sandage69} and \citet{Carney79} normalized the ultra-violet excess by using 
the procedure of \citet{Wildey62}. \citet{Sandage69} compared the $U-B$ colours of 
maximum abundance (which corresponds to $B-V=0.60$) for 112 stars of large proper motion 
with that of Hyades for the same $B-V$ colour, and defined 16 ``guillotin factors'' 
as follows for normalization the ultra-violet excesses: $f=\delta_{0.6}(U-B)/\delta(U-B)$, 
where $\delta_{0.6}(U-B)$ and $\delta(U-B)$ are the ultra-violet excesses at $B-V=0.6$ 
and at a given $B-V$ colour which needs to be normalized. 

Other studies followed the pioneer ones, i.e. \citet{Carney79, Karaali03, 
Karaali11, Karatas06, Guctekin16}. Other researchers calibrated the iron abundance in 
terms of different photometric indices with different procedures: {\citet{Walraven60, 
Stromgren66, Cameron85, Buser90, Trefzger95}. The alpha elements -Mg, Ti, Ca, Na, O, 
Si- became important phenomena in recent years. The researchers plotted these elements 
either individually or their combinations against iron abundances for a set of star 
sample and separated them into different populations, i.e. thin disc, thick disc, 
and halo. All the alpha elements appeared in the literature have been determined by 
their spectroscopic means. However, this procedure requires high resolution which is 
limited with the near-by dwarfs. We thought to extend the determination of the alpha 
elements to larger distances by calibrating them photometrically and it is the main 
scope of this study. 

The first time, we estimated synthetic alpha elements, [$\alpha$/Fe]$_{syn}$, and 
ultra-violet excesses, $\delta_{syn}$ using the Dartmooth Stellar Evolution 
Database\footnote{http://stellar.dartmouth.edu/models/} \citep{Dotter08} and compared 
their distribution with the estimated ones by our calibration. We will see that there 
is an agreement between two sets of data.  

Stars for which the alpha elements are available in the literature were separated into 
different populations in the corresponding studies. While the metallicity calibrations 
are free of populations. Hence, we will concern only with the alpha elements but not 
with population types of the sample stars. We organized the paper as follows: The data 
are given in Section 2 and the procedure is explained in Section 3. Finally, Section 4 
is devoted to a summary and discussion.

\section{Data}
We have two sets of data in our library. The first one consists of the abundances of 
different elements of \citet[][hereafter V04]{Venn04}, \citet[][hereafter B14]{Bensby14}, 
\citet[][hereafter R06]{Reddy06}, \citet[][hereafter N10]{Nissen10}, and \citet[]
[hereafter S02]{Stephens02}. The data of V04 is a collection of the data in 15 studies, 
eight of which are supplied with kinematics while seven of them are without kinematics. 
The abundances of 10 elements are present for 780 stars in V04. V04 separated the stars 
into thin disc (TN), thick disc (TK), and halo (H) populations. The catalogue of B14 
includes the abundances of 12 elements for 714 stars, including the iron element 
[Fe/H]. Kinematical data are also available in this catalogue. R06 measured the abundances 
of 11 elements, including [Fe/H], for 176 stars and they used them to separate the stars 
into different populations as in V04. Kinematic data are also available in this catalogue. 
A similar investigation can be seen in N10 who measured the abundances of eight elements 
including [Fe/H] for 100 stars. However they used a different notation in separation of the 
stars into different populations, i.e. TD (thick disc stars), ``high-$\alpha$'' and 
``low-$\alpha$'' stars. S02 covers the abundances of eight elements including [Fe/H] 
for 56 halo stars. There is a large overlapping of the stars in five studies mentioned above. 
We gave priority to the stars in V04 and B14 and applied a series of constraints to 
obtain the final set of data available for alpha element calibration, as explained 
in the following: we reduced the multiplicity of the stars to a single one, we considered 
only the stars for which the alpha elements [Mg/Fe], [Ca/Fe], [Ti/Fe], and [Na/Fe] 
are available, and we omitted the giants and the stars without $U-B$ and $B-V$ colour 
indices. Thus the number of stars (dwarfs) in this set reduced to 619.

The second set of data is the Hypatia catalogue \citep{Hinkel14}. This catalogue is a 
collection of many elements published in 69 studies. The least number of elements measured 
in these studies is two \citep{Ecuvillon06,Caffau11}, while the biggest one is 33 
\citep{Galeev04}. The number of stars observed in different studies are also different, 
i.e. \cite{Neuforge97} and \cite{Porto08} measured only two stars, while \cite{Petigura11} 
and \cite{Valenti05} measured as much as 914 and 1002 stars, respectively. The total number 
of stars in Hypatia catalogue is 8821. However, there are many overlapping stars in different 
studies (including the ones in the first set mentioned in the first paragraph of this section) 
whose data are included in this catalogue. We applied the same constraints to the data 
in the second set which reduced the number of stars to only 43. 

Then, we decided to separate 39 dwarfs from the first set and combine them with the 43 ones 
in the second set for the application of the calibration, and use the remaining 589 dwarfs 
in the first set for the calibration of the alpha elements. We should emphasize that 
the mentioned 39 stars are not included in the sample used for the calibration. However, 
these stars were observed with the sample stars simultaneously. While, the 43 stars taken 
from the Hypatia catalogue were observed at different times. Thus, we should have a chance 
to compare the results obtained from the two different sets of stars and discuss their accuracy. 
We evaluated the mean of the anti-logarithm values of [Mg/Fe], [Ca/Fe], [Ti/Fe] and 
[Na/Fe] common in all stars and took the logarithm of this mean value  for each star 
which will be called as ``$\alpha$ element relative to iron'', i.e. [$\alpha$/Fe], in this 
study. The authors in the cited studies plotted their alpha elements against the iron elements 
and interpreted this distribution. One can see this picture in all studies related to the 
alpha elements, whereby thin disc, thick disc and halo populations are defined. Our aim is 
different. As stated in the Introduction, one does not need the population types of the 
sample stars in any metallicity calibration. Hence, we used only the alpha elements of 
our sample stars and fitted them to their ultra-violet excesses obtained from the $UBV$ 
data. The $U-B$ and $B-V$ data of these stars are provided from 
SIMBAD\footnote{http://simbad.u-strasbg.fr/simbad/sim-fbasic} and they were de-reddened 
by the procedure as explained in the following. The $E(B-V)$ colour excesses were evaluated 
individually for each star making use of the maps of \cite{Schlafly11}, and it was reduced 
to a value corresponding to the distance of the star by means of the equation of \cite{Bahcall80}. 
Then, the $E(U-B)$ colour excesses were evaluated by the following equation \citep{Gar88};

\begin{eqnarray}
E(U-B)=0.72E(B-V)+0.05E^{2}(B-V).
\end{eqnarray}
Finally, the de-reddened $(B-V)_0$ and $(U-B)_0$ colours could be determined 
by the following equations:
\begin{eqnarray}
(U-B)_0=(U-B)-E(U-B),\\ \nonumber
(B-V)_0=(B-V)-E(B-V).
\end{eqnarray}

The stars cover the $(B-V)_0$ colour range $0.325< (B-V)_0 \leq 0.775$ mag which corresponds 
to the spectral type range F0-K2. The spectroscopic and photometric data of the stars, as 
well as their ID numbers, coordinates, and sources are given in Table 1. A detailed inspection 
of these stars revealed that some of them have some peculiarities, i.e. they are binary, variable, 
double or multiple, and chromospheric active stars. We did not exclude such stars from the program, 
however we followed a procedure, explained in Section 3, to omit those with large scatter in the 
diagram that we used for calibration of the [$\alpha$/Fe] in terms of ultra-violet excess.

\section{The Procedure}
The procedure consists of calibration of the [$\alpha$/Fe] element in terms of ultra-violet 
excess $\delta(U-B)_0$, the difference between $(U-B)_0$ colours of a given star and the Hyades 
star with the same $(B-V)_0$ colour. $\delta(U-B)_0$ excess is related to a bulk heavy metal 
abundance which is denoted as [M/H] in the literature, and [M/H] can be related to other 
indicators for various components of the Milky Way Galaxy. These indicators cover small group of 
elements, where [Fe/H] is the most favorite indicator in question. Thus, as stated in the 
Introduction, [Fe/H] could be calibrated in terms of $\delta(U-B)_0$ and the iron abundances 
of the dwarf stars in our Galaxy could be determined. Here, the indicator will be the [$\alpha$/Fe] 
element which covers the elements [Mg/Fe], [Ca/Fe], [Ti/Fe], [Na/Fe]. Stars with 
different $(B-V)_0$ colours with the same metal abundance show different values of ultra-violet 
excess due to the shapes of the blanketing vectors as stated in the Introduction. For late-type 
stars, $\delta(U-B)_0$ is partially guillotined because the blanketing line is nearly parallel 
to the intrinsic Hyades line. Hence, the ultra-violet excesses of stars used in the calibration 
of any metallicity should be normalized. This is the case in \cite{Sandage70} and \citet{Karaali03, 
Karaali05, Karaali11} where all ultra-violet excesses were reduced to the ultra-violet excess at 
$(B-V)_0=0.6$ mag, $\delta_{0.6}(U-B)_0$. There is a slight different procedure for calibration 
of [$\alpha$/Fe] –and any metallicity– in terms of ({\em unreduced}) ultra-violet excess, i.e. 
one can fit the [$\alpha$/Fe] values to the $\delta(U-B)_0$ ultra-violet excesses for a sample 
of stars with limited $(B-V)_0$ colour-range. This is the procedure we used in this study. 
We separated the colour interval $0.325<(B-V)_0 \leq 0.775$ mag into nine sub-intervals, i.e. 
$0.325<(B-V)_0 \leq 0.375$, $0.375<(B-V)_0 \leq 0.425$, ..., and $0.725<(B-V)_0 \leq 0.775$ mag. 
Thus we obtained nine sub-samples of stars for calibration of [$\alpha$/Fe] in terms of $\delta(U-B)_0$ 
(hereafter we will use the notation $\delta$). The scales of the sub-intervals are equal, 
$\Delta(B-V)_0=0.05$ mag, and they are larger than the errors in the $(U-B)_0$ and $(B-V)_0$ colours.     

\subsection{Calibration of [$\alpha$/Fe] in terms $\delta$}
We separated the stars in Table 1 into nine sub-samples as stated in the preceding paragraph and 
calibrated the [$\alpha$/Fe] elements in terms of ultra-violet excesses $\delta$ for stars in each 
sub-sample as explained in the following. Each calibration is carried out in three steps. In the 
first step, we rejected the stars which show large scatter in the [$\alpha$/Fe] - $\delta$ diagram. 
These stars are candidates for binarity, variable, double or multiple, and chromospheric active 
stars. Then, we fitted the [$\alpha$/Fe] element in terms of $\delta$ (first calibration) for 
the remaining sample stars and reproduced the alpha elements of the stars in this reduced sample 
by replacing their ultra-violet excesses into the first calibration. We use the symbol 
$[\alpha/Fe]_{rep}$ for the alpha elements reproduced by the first calibration. Then, we estimated 
the residuals, the differences between the (original) [$\alpha$/Fe] elements and [$\alpha$/Fe]$_{rep}$, 
and the corresponding standard deviation, $\sigma$. 

In the second step we omitted stars with (original) [$\alpha$/Fe] elements which lie out of the 
interval [$\alpha$/Fe]$_{rep} \pm \Delta$[$\alpha$/Fe], where $\Delta$[$\alpha$/Fe] corresponds to 
the mean of the residuals larger than the standard deviation, $\sigma$. Thus, we re-reduced 
the original star sample for the second time and (in the third step) we calibrated the (original) 
[$\alpha$/Fe] element in terms of $\delta$. This is our second and final calibration for a sub-sample. 
The results are given in Fig. 1. The stars rejected in the first step and omitted in the second step are 
shown by an asterisk and a triangle, respectively, while filled circles indicate the stars used in the 
final calibration.

\begin{figure*}
\begin{center}
\includegraphics[scale=0.8, angle=0]{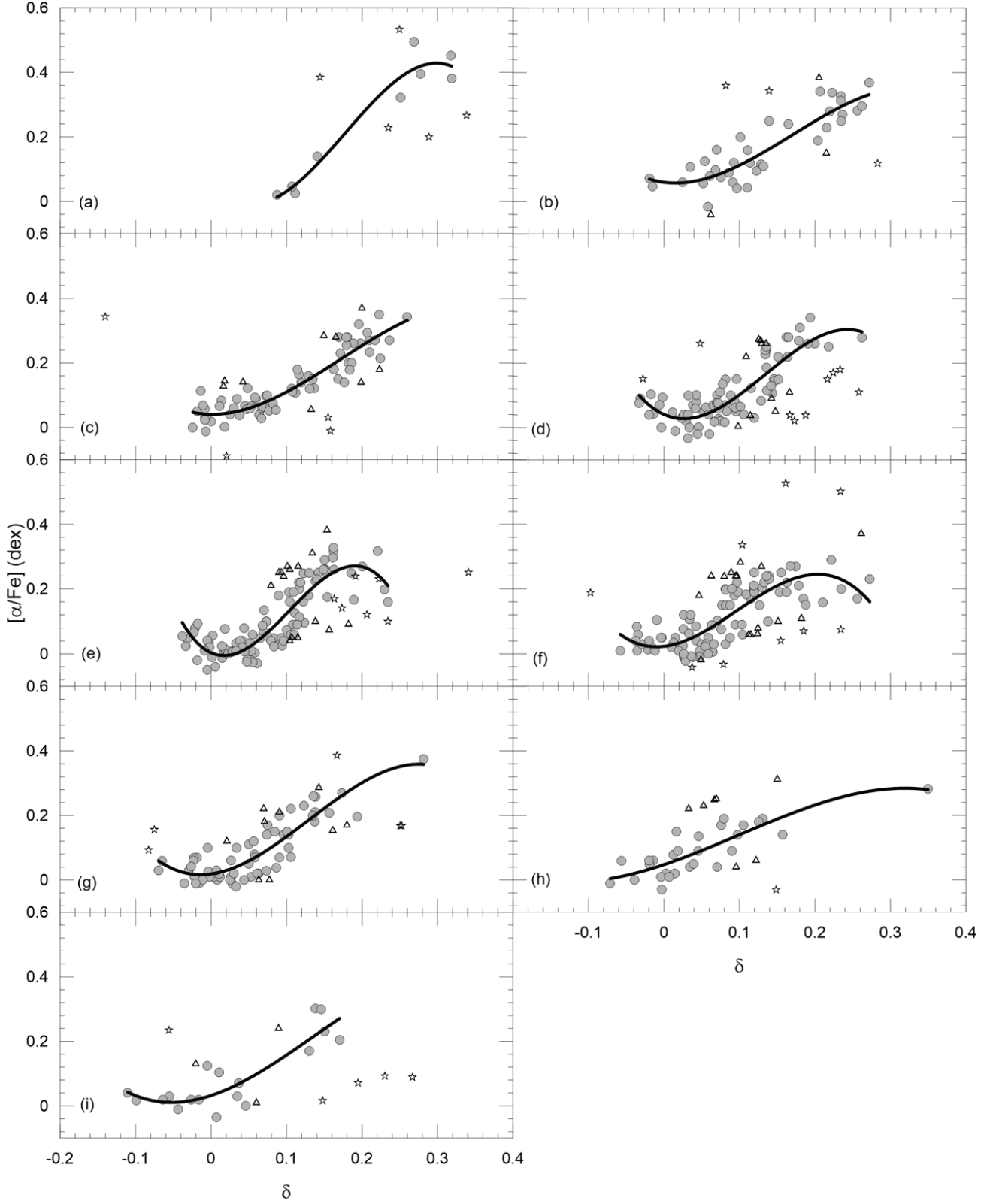}
\caption[]{Calibration of the [$\alpha$/Fe] element in terms of $\delta$ for nine sub-samples: (a) $0.325<(B-V)_0 \leq 0.375$, (b) $0.375<(B-V)_0 \leq 0.425$, (c) $0.425<(B-V)_0 \leq 0.475$, (d) $0.475<(B-V)_0 \leq 0.525$, (e) $0.525<(B-V)_0 \leq 0.575$, (f) $0.575<(B-V)_0 \leq 0.625$, (g) $0.625<(B-V)_0 \leq 0.675$, (h) $0.675<(B-V)_0 \leq 0.725$, (i) $0.725<(B-V)_0 \leq 0.775$ mag. An asterisk indicates the star with large scattering which is rejected from the program (first step in the text), a triangle indicates the star which lies out of the interval [$\alpha$/Fe]$_{rep} \pm \Delta$ [$\alpha$/Fe] and which is omitted in the final calibration (second step in the text), and a filled circle indicates the star considered in the final calibration (third step in the text). The symbols $[\alpha/Fe]_{rep}$ and $\Delta$ [$\alpha$/Fe] are explained in the text.} 
\end{center}
\end {figure*}

The total number of stars used in our (final) calibrations is 469. We adopted a third degree 
polynomial in our calibrations as in the following: 

\begin{eqnarray}
[\alpha/Fe]=a_3\delta^3+a_2\delta^2+a_1\delta^1+a_0.
\end{eqnarray}
The numerical values of the coefficients $a_i$ ($i=0, 1, 2, 3$) for all the stars except the rejected ones 
are given in Table 2, while those for the re-reduced sub-sample (the final one without rejected and omitted 
stars) are presented in Table 3. The coefficients which will be used for the estimation of the alpha 
elements via the ultra-violet excess are the ones in Table 3. 

\subsection{Application of the Procedure}
We applied the procedure to two sets of field stars and two sets of clusters. 
The first set of field stars consists of 43 stars taken from the Hypatia catalogue 
\citep{Hinkel14}, while the second one is a combination of stars in \citet[][hereafter Bai04]{Bai04}:10, 
B14: 12, V04: 7, S02: 5, N10: 3, and R06: 2, totally 39 stars. The difference between these sets of stars 
is that 29 stars in B14, V04, S07, N10 and R06 are taken from the same source of stars used for the 
calibration of the [$\alpha$/Fe] element. However, they are not included into the sample used for the 
calibration, as stated in Section 2. We evaluated the ultra-violet excesses ($\delta$) of the stars in 
both sets by using their $UBV$ data provided from SIMBAD and the procedure in Section 2, and replaced 
them into our final calibration to calculate an alpha element, [$\alpha$/Fe]$_{cal}$, for each star. Now, 
we need the original alpha elements, [$\alpha$/Fe]$_{org}$, of the stars in question which are based on 
spectroscopic means to test the accuracy of the calculated alpha elements. We evaluated the mean of the 
anti-logarithmic values of the [Mg/Fe], [Ca/Fe], [Ti/Fe] and [Na/Fe] elements common in the stars for 
each set and took the logarithm of the mean value for each star for our purpose.

The results for the first set of the field stars taken from the Hypatia catalogue are given 
in Table 4a. The mean of the residuals, $\Delta$[Fe/H]=[$\alpha$/Fe]$_{org}$-[$\alpha$/Fe]$_{cal}$, 
and the corresponding standard deviation are $\langle \Delta$[$\alpha$/Fe]$\rangle$=-0.090 and 
$\sigma=0.102$ dex, respectively. The results for the second set of the field stars, taken from the same 
source of the stars used for the calibration of the [$\alpha$/Fe], are tabulated in Table 4b. The mean 
of the residuals and the corresponding standard deviation are $\langle \Delta$[$\alpha$/Fe]$\rangle$=-0.009 
and $\sigma=0.079$ dex, respectively. One can see that the mean of the residuals is rather small, and the 
standard deviation is smaller than the one 43 field star.

\cite{Dias16} measured the iron abundance [Fe/H] and alpha element [$\alpha$/Fe] for 51 globular 
clusters in their FORS2/VLT survey. We used this advantage to apply our calibrations to two sets 
of globular clusters as explained in the following. We could provided the $B-V$ and $U-B$ colour 
indices for 25 of these clusters from the catalogue of \cite{Harris10}, and for 19 clusters from 
\cite{Hanes85}. The restriction of the number of clusters is due to the constraint of the range 
of the $(B -V)_0$ colour index of our calibration, i.e. $0.325<(B-V)_0 \leq 0.775$ mag. The 
$B-V$ and $U-B$ colour indices for 16 clusters are available in both sets. However, the values 
of a given colour index for a cluster are different in two sets which cause different residuals as it is 
shown in the following.

The data for the 25 clusters of the first set whose $B-V$ and $U-B$ colour indices are taken 
from \cite{Harris10} are given in Table 4c. The de-reddened colours are evaluated by using the 
colour excess $E(B-V)$ in \cite{Harris10}, the ultra-violet index for the Hyades cluster, $(U-B)_H$, 
the ultra-violet excess, $\delta$, and the [$\alpha$/Fe]$_{cal}$ are estimated by the same procedures 
used in the preceding paragraphs. The mean of the residuals, $\Delta$[Fe/H]=[$\alpha$/Fe]$_{org}$-[$\alpha$/Fe]$_{cal}$, 
and the corresponding standard deviation are $\langle \Delta$[$\alpha$/Fe]$\rangle$=0.073 and $\sigma=0.091$ dex.  

The data for 19 clusters for the second set are presented in Table 4d. The errors for $B-V$ of 
three clusters and for $U-B$ of 11 clusters are also given in \cite{Hanes85}. We considered the errors tabulated 
in Table 4d for six clusters to obtain  smaller residuals in [$\alpha$/Fe] for them, while for five clusters the 
residuals are already small and one does not need any correction in $B-V$ or $U-B$ colour indices. The mean of 
the residuals and the corresponding standard deviation 
for the second set of the clusters are much better than for the first set, 
i.e. $\langle \Delta$[$\alpha$/Fe]$\rangle$=-0.012 and $\sigma = 0.053$ dex, respectively. 

\subsection{Synthetic Alpha Elements and UV-Excesses}
We estimated synthetic alpha elements, [$\alpha$/Fe]$_{syn}$, and ultra-violet excesses, $\delta_{syn}$, for nine 
sub-samples, i.e. $0.325<(B-V)_0\leq 0.375$, $0.375<(B-V)_0\leq 0.425$, ..., $0.725 <(B-V)_0\leq 0.775$ mag, using 
the Dartmouth Stellar Evolution Program \citep[DSEP,][]{Dotter08} and compared their distribution with the ones 
estimated by our calibration, as explained in the following. We took the DSEP isochrones and did the necessary 
interpolations to get the required data for the relations.  The estimations of [$\alpha$/Fe]$_{syn}$ and 
$\delta_{syn}$ are carried out for three populations, thin disc, thick disc and halo whose metallicity ranges 
are assumed to be -0.5$<$[Fe/H]$\leq$ +0.5, -1.0$\leq$ [Fe/H]$<$-0.5 and -2.5$\leq$ [Fe/H]$<$-1.0 dex, respectively. 
Also, we adopted the 3, 12 and 13 Gyr as the ages of the populations in the same order. Thus, we estimated a set 
of $(B-V, U-B)$ couples for each population using an iron abundance, an [$\alpha$/Fe]$_{syn}$ value and an age value, 
each time. The range of the [$\alpha$/Fe]$_{syn}$ is adopted as -0.2$\leq$[$\alpha$/Fe]$_{syn}\leq$ +0.8. The iron 
metallicity and [$\alpha$/Fe]$_{syn}$ are used in 0.5 and 0.2 dex steps, respectively. Finally, we considered only 
the $B-V$ and $U-B$ colour indices  which lie in the $(B-V)_0$ range of our sample, i.e. $0.325<(B-V)_0\leq 0.775$ dex. 
The ultra-violet excesses are estimated relative to the ultra-violet index for [Fe/H]=0.0 dex. The distributions of 
the [$\alpha$/Fe]$_{syn}$ and $\delta_{syn}$, are plotted in Fig. 2. The mean $(B-V)_0$ colour index and a colour coded 
symbol are also indicated for each sub-sample. We could fit the data in Fig. 2 to a third degree polynomial with a high 
correlation coefficient ($R^2=0.987$): 
[$\alpha$/Fe]$_{syn}=-22.827\times \delta_{syn}^3+10.604\times\delta_{syn}^2+0.264\times \delta_{syn}+0.017$. Finally, 
we plotted our calibrations obtained for nine sub-samples and the one for synthetic data in the same diagram for 
comparison purpose. Fig. 3 shows that the calibration curves and the synthetic one have the same trend. Also, for 
$\delta_{syn}<0.18$ the synthetic curve lies within the region occupied by the calibration curves. Although the 
synthetic curve for larger $\delta_{syn}$ occupies the alpha elements which are a bit larger than the ones 
corresponding to the calibration curves, an additional error of $\Delta$[$\alpha$/Fe]=0.10 dex to our calibrations 
supplies an agreement also for this segment. The agreement of the synthetic calibration with the ones based on the 
measured alpha elements indicates that the alpha abundances of individual stars can be determined via their UV excesses. 

\begin{figure*}
\begin{center}
\includegraphics[scale=0.85, angle=0]{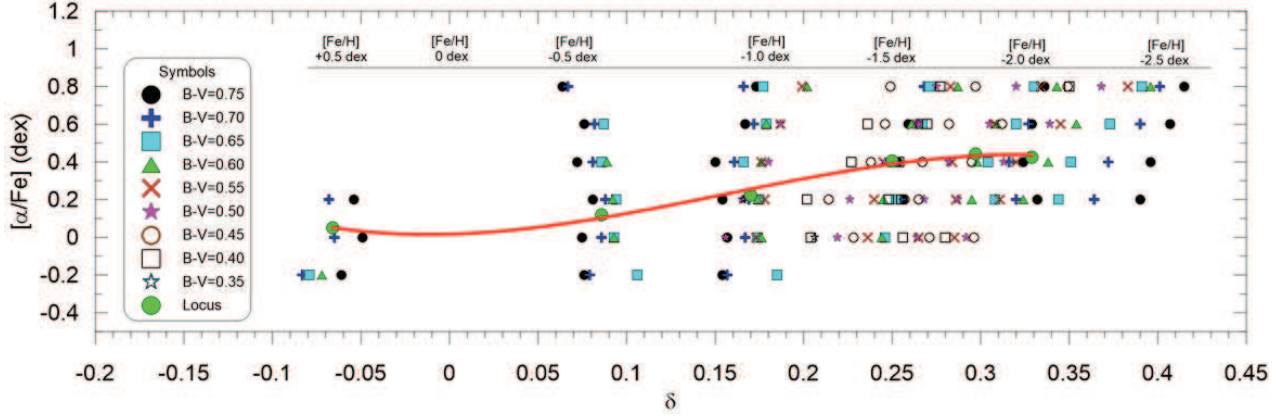}
\caption[]{([$\alpha$/Fe]$_{syn}$, $\delta_{syn}$) diagram for a set of iron metallicity 
which covers thin disc, thick disc, and halo populations.The curve indicates the third degree 
polynomial fitted to the data. The range of the synthetic alpha element is 
-0.2$\leq$[$\alpha$/Fe]$_{syn}\leq$ +0.8 dex. The colour coded symbols are explained in the text.} 
\end{center}
\end {figure*}

\begin{figure*}
\begin{center}
\includegraphics[scale=0.85, angle=0]{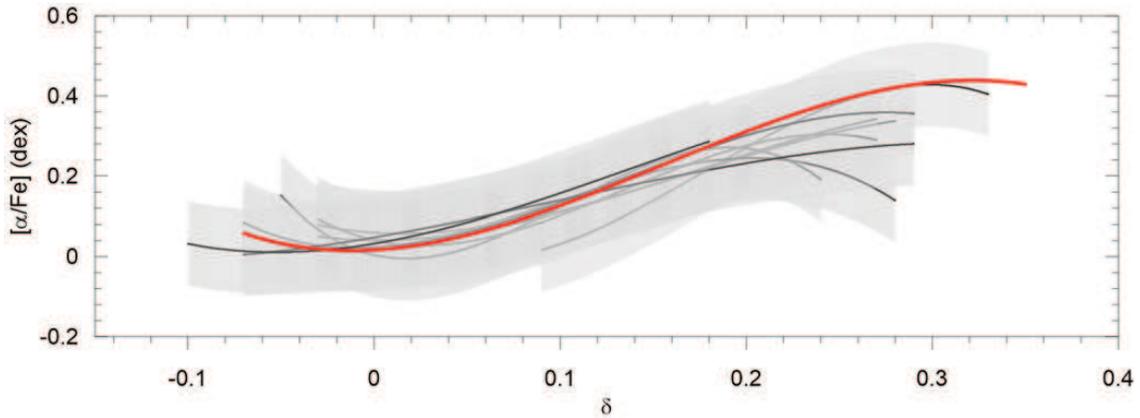}
\caption[]{Comparison of the calibration based on the synthetic alpha elements 
and ultra-violet excesses, [$\alpha$/Fe]$_{syn}$ and $\delta_{syn}$ (red curve), with the calibrations 
obtained for nine sub-samples. The shadowed area corresponds to nine calibrations with 0.10 uncertainty 
in [$\alpha$/Fe]$_{syn}$.} 
\end{center}
\end {figure*}

\section{Summary and Discussion}

We present the calibration of the [$\alpha$/Fe] element in terms of the ultra-violet excess, $\delta$, for dwarf stars 
in nine sub-samples, defined by the $(B-V)_0$ colours with a range of $\Delta(B-V)_0=0.050$, i.e. $(0.325, 0.375]$, 
$(0.375, 0.425]$, ..., and $(0.725, 0.775]$. We fitted a third degree polynomial for our calibration, a procedure that 
we used in \citet{Karaali03, Karaali05, Karaali11} for calibration of the iron abundance, [Fe/H]. The difference 
between our study and the cited ones is that in the previous studies all the ultra-violet excesses were normalized to 
the colour $(B-V)_0=0.6$, and only a single calibration equation was derived for all the data. Whereas, here we 
preferred to use the ultra-violet excess estimated for the colour of the star. However, we separated the stars into 
sub-samples with limited $(B-V)_0$ range to avoid any guillotin effect mentioned in Section 3. The lower limit of the 
[$\alpha$/Fe] element in our calibrations is $\sim 0$ for all $(B-V)_0$ intervals, while the upper limits for the blue 
colours $(0.325, 0.375]$, $(0.375, 0.425]$, and $[0.425, 0.475]$ are larger than the ones for redder colours. Our 
calibrations provide [$\alpha$/Fe] elements in the range (0.0, 0.4], which cover the elements of the thin disc, 
thick disc and some halo stars. 

We applied the procedure to two sets of field stars and two sets of clusters. The 
mean of the residuals and the corresponding standard deviation for 43 field stars taken 
from Hypatia catalogue \citep{Hinkel14} are $\langle \Delta$[$\alpha$/Fe]$\rangle$=-0.090 
and $\sigma = 0.102$ dex, while those for the second set of 39 field stars which are taken 
from Bai04, B14, V04, S02, N10 and R06 are $\langle \Delta$[$\alpha$/Fe]$\rangle$=-0.009 and 
$\sigma = 0.079$ dex. We compared the numerical values of alpha elements estimated for a star 
by different researchers to test the accuracy of our results. The data for [Mg/Fe], [Ca/Fe], 
and [Ti/Fe] elements in Table 5a are taken from Bai04, while their mean, [$\alpha$/Fe], is estimated 
in this study. One can see that the differences between the numerical values for the cited elements 
determined by Bai04, F00, and S02 can be as large as 0.20 and 0.30 dex. There are also differences 
between the numerical values of [$\alpha$/Fe] elements determined by V04, B14, N10, and R06, 
however not larger than $\sim 0.10$ dex (Table 5b). It seems that the accuracy of the [$\alpha$/Fe]$_{cal}$ elements 
for the 43 Hypatia stars is compatible to the one for the alpha elements in Table 5a,while the [$\alpha$/Fe]$_{cal}$ 
elements estimated for the 39 stars in the second set are more accurate and their accuracy is compatible to the alpha 
elements in Table 5b. 39 stars with more accurate [$\alpha$/Fe]$_{cal}$ elements were measured with the sample stars 
simultaneously which indicates (a small) bias, i.e. the accuracy of the alpha elements depends also on the procedure 
used in their observations.

\begin{figure}
\begin{center}
\includegraphics[scale=0.85, angle=0]{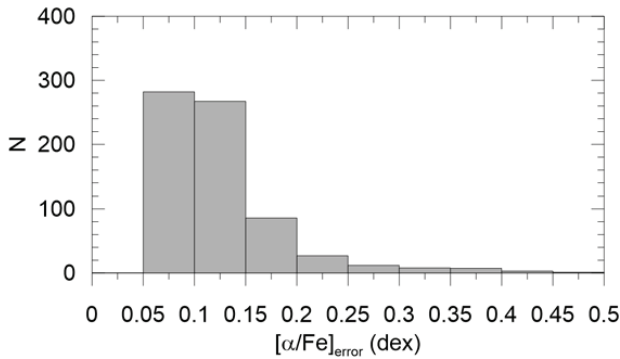}
\caption[]{Distribution of the propagated errors for the [$\alpha$/Fe] element for the stars in B14.} 
\end{center}
\end {figure}

We applied the procedure also to two sets of globular clusters whose [$\alpha$/Fe] elements 
are taken from the same study \citep{Dias16}. \cite{Harris10} catalogue provides $B-V$ and 
$U-B$ colour indices for 25 clusters for the application of our calibration, while the 
number of clusters in \cite{Hanes85} with different $B-V$ and $U-B$ colours available 
for our purpose is 19. The colour excesses, $E(B-V)$, for all clusters are taken from \cite{Harris10}. 
We expect from this procedure to find out the reason of any probable difference 
in accuracy for the estimated alpha elements in two sets. The mean of the residuals and 
the corresponding standard deviation for 25 clusters in the first set are 
$\langle\Delta$[$\alpha$/Fe]$\rangle$=0.073 and $\sigma = 0.091$ dex, while the ones for 19 clusters 
in the second set are $\langle \Delta$[$\alpha$/Fe]$\rangle$=-0.012 and $\sigma = 0.053$ dex, respectively. 
One can see that there are considerable differences between the mean of the residuals, and standard deviations 
for two set of clusters. We used the same calibrations to estimate the [$\alpha$/Fe]$_{cal}$. 
Hence the differences in question should originate from the data used in two sets. 
Original alpha elements, [$\alpha$/Fe], are taken from the same study. Additionally there are 
16 clusters common in two sets which means that 16 residuals in two sets are estimated 
by the same [$\alpha$/Fe] elements. All the colour excesses, $E(B-V)$, are taken from the 
same source, and the ultra-violet index for the Hyades cluster, $(U-B)_H$, is estimated by 
the same equation. Then, it remains the $B-V$ and $U-B$ colour indices which cause different 
residuals and standard deviations. The residuals, $\Delta$[$\alpha$/Fe], for 16 clusters in 
the first set (Table 4c) are added to the last column of Table 4d for comparison purpose. 
One can see that different $B-V$ and $U-B$ colour indices of a star may cause a difference 
in the residual as much as 0.18 dex. One can generalize this result obtained via the clusters, 
i.e. the accuracy of an estimated alpha element for a star depends on the accuracy of the 
 $B-V$ and $U-B$ colour indices of that star. 

\begin{figure}
\begin{center}
\includegraphics[scale=0.9, angle=0]{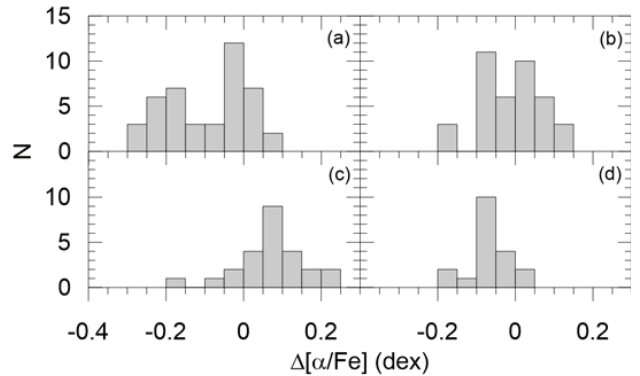}
\caption[]{Histograms of the residuals, $\Delta$[$\alpha$/Fe], in Tables 4a, 4b, 4c and 4d.} 
\end{center}
\end {figure}

As stated in Section 2, most of the stars used for the calibration of the procedure are taken 
from B14 (238 stars) and V04 (283 stars). The errors for the elements [Mg/Fe], [Ca/Fe], 
[Ti/Fe] and [Na/Fe] were not given in V04, while they are present in B14. We used them 
and estimated the propagated error for the alpha element, [$\alpha$/Fe], for all stars 
(714 stars) in B14 and showed their distribution in Fig. 4. The mean and the standard deviation 
for the errors in question are $\langle$ [$\alpha$/Fe]$_{err}\rangle$= 0.147 and $\sigma = 0.094$ 
dex, respectively. However, the propagated errors for the majority of the stars are less than
 0.15 dex.

We thought that it would be useful to plot the differences between the original and 
estimated alpha elements in a figure. This is carried out in  Fig. 5, where panels 
(a), (b), (c) and (d) cover the distributions of the residuals, 
$\Delta$[Fe/H]=[$\alpha$/Fe]$_{org}$-[$\alpha$/Fe]$_{cal}$, in Tables 4a, 4b, 4c, and 4d, respectively. 
One can see that the absolute values of the residuals for the majority of the stars 
in four panels are less than 0.15 dex, a numerical value equals to the maximum propagated error 
for the original alpha elements for most of the stars in B14. Statistics of the
residuals (absolutely) less than 0.15 as well as 0.20 dex are given in Table 6. 
We estimated synthetic alpha elements and UV - excesses too. There is an agreement 
between the synthetic calibration and the observed ones within $\Delta$[$\alpha$/Fe]=0.10 dex accuracy.

{\bf Conclusion:} The accuracy of the alpha elements estimated in our calibration and the 
agreement of the synthetic calibration with our calibration based on the measured alpha elements 
confirm our argument that alpha abundances of individual stars based on the UV excess of stars 
can be constrained. 

\section*{Acknowledgments} 
Authors are grateful to the anonymous referee for his/her considerable contributions to improve 
the paper. This research has made use of NASA (National Aeronautics and Space Administration)'s 
Astrophysics Data System and the SIMBAD Astronomical Database, operated at CDS, Strasbourg, France.

\begin{landscape}
\textwidth = 700 pt
\begin{table*}
\setlength{\tabcolsep}{0.5pt}
\tiny{
\caption{Data for 589 program stars. The columns give: current number ($N$), ID, equatorial coordinates ($\alpha,\delta$), Galactic coordinates ($l$, $b$), parallax ($\pi$) and its error ($\pi_{err}$), colour excess ($E(B-V)$), $B-V$ and $U-B$ colours, de-reddened $(B-V)_0$ and $(U-B)_0$, effective temperature ($T_{e}$), surface gravity ($\log g$), iron abundance ([Fe/H]), alpha element ([$\alpha$/Fe]), remarks, ultra-violet colour corresponding Hyades squence, $(U-B)_H$, for a given $(B-V)_0$ colour, ultra-violet excess ($\delta$), and Remark}
\begin{tabular}{lcccccccccccccccccccccc}
    \hline
$N$ &ID	   &	$\alpha$      & $\delta$    &     $l$  &   $b$	&$\pi$ & $\pi_{err}$ & $E(B-V)_{r}$ & $B-V$&  $U-B$ & $(U-B)_0$ & $(B-V)_0$& $T_e$ & $\log g$ & [Fe/H] & [$\alpha$/Fe]& Ref.   &Rem.& Ref &$(U-B)_H$&$\delta$&Remark\\
    &  	   &    (hh:mm:ss.ss) &(dd:mm:ss.s) &  ($^{o}$)& ($^{o}$)    &(mas) & (mas)       &   (mag)      & (mag)&  (mag) &  (mag)    & (mag)    & (K)   &   (cgs)  &   (dex)  &   (dex)      &(dex)   &    &     &  (mag)  & (mag)&\\
\hline
1  &HIP 36513   &07:30:41.30	&$+$24:05:10.0 &194.9332 &18.939	&2.57 &2.32		&0.031	    &0.357&-0.220 &-0.242   &0.326    & $-$  &$-$ & -2.55&0.38&V04& HP  & $-$  &0.077&0.32& considered\\
2  &HIP 96115   &19:32:31.90	&$+$26:23:26.0 & 60.9206 &3.4682	&7.25 &1.35		&0.054	    &0.389&-0.218 &-0.257   &0.335    & $-$  &$-$ & -2.41&0.45&V04& HP  & $-$  &0.061&0.32& considered\\
3  &HIP 47048   &09:35:16.70	&$-$49:07:48.9 &273.2968 &2.117		 &10.26&0.95		&0.032	    &0.398&-0.095 &-0.118   &0.366    &6630  &4.12& -0.50&0.14&B14&$-$  &$-$   &0.023&0.14& considered\\
  .&           .&              .&        .     &             .&.         &    .&             .&          .&    .&      .&        .&        .&     .&   .&     .&   .&  .&  .  &     .&    .&     .& .\\
  .&           .&              .&        .     &             .&.         &    .&             .&          .&    .&      .&        .&        .&     .&   .&     .&   .&  .&  .  &     .&    .&     .& .\\
  .&           .&              .&        .     &             .&.         &    .&             .&          .&    .&      .&        .&        .&     .&   .&     .&   .&  .&  .  &     .&    .&     .& .\\
588&HIP 68101   &13:56:32.90    &$-$54:42:17.0 &312.3222 & 6.9756	&25.84&	0.48	&0.012	    &0.780&  0.230&0.222    &0.768    &	$-$  & $-$&-0.53 &0.02&V04& HP  & $-$  &0.37 &0.15        &scattered\\
589&HIP 50713   &10:21:16.79    &$-$17:02:55.4 &259.0917 & 32.7519       &16.48&	1.14	&0.014	    &0.786&  0.327&0.317    &0.772    &	5263 &4.46&-0.11 &0.01&B14&CA&Pace+2013&0.38 &0.06&omitted \\
\hline
\end{tabular}
}
\end{table*}
\end{landscape}


\begin{table*}[]
\centering
\caption{Numerical values for the coefficients $a_i$ ($i=0, 1, 2, 3$) used for the calibration 
of $[\alpha/Fe]$ element in the first step. $N$ indicates the number of stars in each $(B-V)_0$ 
colour range, $\sigma$ is the standard deviation and $R^2$ is the squared correlation coefficient.}
\label{first}
\begin{tabular}{cccccccc}
\hline
$(B-V)_0$ & $a_3$   & $a_2$  & $a_1$   & $a_0$  & $R^2$  & $\sigma$ & $N$    \\ 
\hline
$(0.325, 0.375]$ & -60.955 & 32.423 & -3.0600 & 0.0739 & 0.9483 & 0.044 & 9   \\
$(0.375, 0.425]$ & -27.357 & 12.810 & -0.4762 & 0.0550 & 0.7470 & 0.057 & 39  \\
$(0.425, 0.475]$ & -17.539 & 8.6445 & -0.0163 & 0.0484 & 0.7569 & 0.048 & 75  \\
$(0.475, 0.525]$ & -52.911 & 21.120 & -0.9557 & 0.0406 & 0.6345 & 0.056 & 86  \\
$(0.525, 0.575]$ & -99.929 & 29.822 & -0.7230 & 0.0062 & 0.6290 & 0.065 & 106 \\
$(0.575, 0.625]$ & -26.470 & 7.5664 &  0.5794 & 0.0331 & 0.5142 & 0.065 & 102 \\
$(0.625, 0.675]$ & -23.075 & 8.9882 &  0.4327 & 0.0251 & 0.6697 & 0.056 & 68  \\
$(0.675, 0.725]$ & -0.7373 &-0.9143 &  1.0304 & 0.0636 & 0.4572 & 0.067 & 35  \\
$(0.725, 0.775]$ & -9.9681 & 5.5567 &  0.7237 & 0.0411 & 0.6725 & 0.060 & 21\\ 
\hline
\end{tabular}
\end{table*}

\begin{table*}[6]
\setlength{\tabcolsep}{3.5pt}
{\small
\centering
\caption{Numerical values for the coefficients $a_i$ ($i = 0, 1, 2, 3$) used in the final calibration of [$\alpha$/Fe] element. $N$ indicates the number of stars in each $(B-V)_0$ colour range, $\sigma$ is the standard deviation and $R^2$ is the squared correlation coefficient. The domain of the [$\alpha$/Fe] is also given in the second column. Errors of coefficients are given in parentheses.}
\label{my-label}
\begin{tabular}{ccccccccc}
\hline
$(B-V)_0$        & $\delta$       & $a_3$   & $a_2$  &  $a_1$  & $a_0$  & $R^2$  &$\sigma$& $N$  \\
\hline
(0.325, 0.375] & [ 0.087, 0.319] & -60.9550(60.7871) & 32.4230(39.1548) & -3.0600(7.8686) & 0.0739(0.4694) & 0.9483 & 0.044 & 9  \\
(0.375, 0.425] & [-0.019, 0.272] & -20.3210(17.1027) & 10.2980( 6.3506) & -0.2996(0.6472) & 0.0600(0.0233) & 0.7980 & 0.048 & 36 \\
(0.425, 0.475] & [-0.024, 0.260] & -17.3710(10.7474) &  8.9305( 3.5685) & -0.0272(0.3162) & 0.0407(0.0085) & 0.8601 & 0.036 & 66 \\
(0.475, 0.525] & [-0.033, 0.262] & -54.1660(11.9521) & 21.8760( 3.8644) & -1.0355(0.3400) & 0.0410(0.0104) & 0.7478 & 0.045 & 76 \\
(0.525, 0.575] & [-0.038, 0.234] &-105.8600(13.6112) & 33.0280( 4.0198) & -1.0470(0.3032) & 0.0038(0.0087) & 0.7883 & 0.047 & 91 \\
(0.575, 0.625] & [-0.058, 0.273] & -45.7050( 9.3323) & 13.2730( 2.9961) &  0.2801(0.2394) & 0.0233(0.0089) & 0.6809 & 0.050 & 86 \\
(0.625, 0.675] & [-0.070, 0.282] & -28.3440( 8.2069) & 11.1100( 2.6058) &  0.3240(0.1900) & 0.0190(0.0083) & 0.7659 & 0.044 & 59 \\
(0.675, 0.725] & [-0.071, 0.350] &  ~-6.8017( 7.8702)&  2.0239( 3.1811) &  0.7881(0.2338) & 0.0477(0.0113) & 0.6888 & 0.042 & 28 \\
(0.725, 0.775] & [-0.110, 0.170] & -14.9820(32.1468) &  6.2218( 3.0120) &  0.7757(0.4553) & 0.0324(0.0189) & 0.7721 & 0.051 & 18 \\
\hline 
\end{tabular}
}
\end{table*}

\begin{table*}[ht]
 \renewcommand{\thetable}{\arabic{table}a}
\setlength{\tabcolsep}{2.5pt}
\center
\fontsize{8pt}{8pt}
\selectfont
 \caption{Data for 43 field stars taken from the Hypatia catalogue and used for the application of the procedure. The columns give: Hipparcos number, galactic coordinates ($l$, $b$), parallax ($\pi$), reduced colour excess ($E_d(B-V)$), iron metallicity ([Fe/H]), de-reddened colours ($(B-V)_0$ and $(U-B)_0$), ultra-violet index for the Hyades cluster ($(U-B)_{H}$), ultra-violet excess ($\delta$), original alpha element ([$\alpha$/Fe]$_{org}$), calculated alpha element 
([$\alpha$/Fe]$_{cal}$), and the residual ($\Delta$[$\alpha$/Fe]).}
\begin{tabular}{ccccccccccccccccc}
\hline
HIP    & $l$  & $b$ &  $\pi$& $E_d(B-V)$ & [Fe/H] & $(B-V)_0$ & $(U-B)_0$ & $(U-B)_H$  & $\delta$  & [$\alpha$/Fe]$_{org}$ & [$\alpha$/Fe]$_{cal}$ & $\Delta$[$\alpha$/Fe]   \\
& ($^o$)  & ($^o$) &  (mas)& (mag) & (dex) &  (mag) & (mag) & (mag)  & (mag)  & (dex) & (dex) & (dex)   \\
\hline
1382   & 323.11 & -71.91 & 27.51$\pm$0.64   & 0.001   &  -0.06 & 0.652  & 0.089  & 0.190  & 0.101  & -0.14          & 0.14          & -0.28       \\ 
6744   & 231.77 & -81.87 & 23.27$\pm$0.92   & 0.004   &  0.12  &  0.731  & 0.237  & 0.306  & 0.069  & -0.11          & 0.11          & -0.22       \\ 
6762   & 141.97 & -61.71 & 34.07$\pm$1.00   & 0.004   &   $-$  & 0.734  & 0.302  & 0.311  & 0.009  & -0.13          & 0.04          & -0.17       \\ 
7404   & 161.9  & -72.83 & 26.09$\pm$0.60   & 0.004   &  0.30  & 0.596  & 0.087  & 0.123  & 0.036  & -0.17          & 0.05          & -0.22       \\ 
9404   & 175.06 & -68.40 & 23.28$\pm$1.13   & 0.005   &  0.00  & 0.745  & 0.286  & 0.330  & 0.044  & -0.11          & 0.08          & -0.19       \\ 
16012  & 228.21 & -56.11 & 34.28$\pm$0.88   & 0.002   &  -0.05 & 0.714  & 0.218  & 0.279  & 0.061  & -0.14          & 0.10          & -0.24       \\ 
16738  & 174.32 & -34.65 &  8.92$\pm$0.91   & 0.156   &  0.37  & 0.424  & 0.046  & 0.002  &-0.044  & 0.15           & 0.10          & +0.05       \\ 
17364  & 260.75 & -49.99 & 26.88$\pm$0.46   & 0.003   &  -0.11 & 0.515  & -0.082 & 0.045  & 0.127  & -0.11          & 0.15          & -0.26       \\ 
19781  & 179.01 & -25.45 & 21.62$\pm$1.10   & 0.072   &   $-$  & 0.626  & 0.190  & 0.158  & -0.032 & -0.02          & 0.02          & -0.04       \\ 
20557  & 174.83 & -19.05 & 22.59$\pm$0.77   & 0.051   &  0.12  & 0.469  & 0.001  & 0.015  & 0.014  & -0.10          & 0.04          & -0.14       \\ 
21008  & 177.33 & -19.20 & 21.23$\pm$0.61   & 0.056   &  0.10  & 0.411  & -0.024 & 0.002  & 0.026  & -0.01          & 0.06          & -0.07       \\ 
22449  & 191.45 & -23.07 & 23.94$\pm$0.17   & 0.012   &  0.05  & 0.439  & -0.018 & 0.004  & 0.022  & 0.01           & 0.04          & -0.03       \\ 
24829  & 257.46 & -34.98 & 28.00$\pm$0.23   & 0.004   &  0.13  & 0.509  & 0.007  & 0.041  & 0.034  & -0.01          & 0.03          & -0.04       \\ 
26617  & 200.92 & -14.06 & 08.79$\pm$2.67   & 0.076   &   $-$  & 0.554  & -0.025 & 0.080  & 0.105  & 0.16           & 0.13          & +0.03       \\ 
29248  & 192.32 & -0.68  & 21.01$\pm$1.12   & 0.005   &  -0.17 & 0.665  & 0.116  & 0.208  & 0.092  & 0.14           & 0.12          & +0.02       \\ 
32480  & 172.36 & 17.52  & 59.82$\pm$0.30   & 0.004   &  $-$   & 0.550  & 0.055  & 0.076  & 0.021  & -0.05          & -0.01         & -0.04       \\ 
33229  & 265.16 & -21.54 & 28.71$\pm$0.51   & 0.008   &  0.14  & 0.775  & 0.357  & 0.383  & 0.026  & -0.11          & 0.06          & -0.17       \\ 
33277  & 190.42 & 12.06  & 58.00$\pm$0.41   & 0.002   &  $-$   & 0.568  & 0.023  & 0.093  & 0.070  & 0.03           & 0.06          & -0.03       \\ 
39903  & 274.97 & -14.95 & 50.05$\pm$2.65   & 0.006   &  $-$   & 0.430  & -0.046 & 0.002  & 0.048  & 0.09           & 0.06          & +0.03       \\ 
40035  & 234.56 & 10.61  & 44.68$\pm$0.30   & 0.002   &  -0.05 & 0.486  & -0.021 & 0.025  & 0.046  & 0.03           & 0.03          &  0.00       \\ 
42173  & 219.12 & 26.20  & 38.11$\pm$0.85   & 0.003   &  0.29  & 0.707  & 0.268  & 0.268  & 0.000  & 0.02           & 0.05          & -0.03       \\ 
46076  & 209.36 & 42.33  & 30.20$\pm$1.03   & 0.006   &  0.00  & 0.634  & 0.136  & 0.167  & 0.031  & -0.05          & 0.04          & -0.09       \\ 
50473  & 227.30 & 51.40  & 30.90$\pm$0.68   & 0.007   &  -0.08 & 0.553  & 0.055  & 0.079  & 0.024  & 0.01           & 0.00          & +0.01       \\ 
51415  & 292.16 & -11.71 & 16.71$\pm$0.89   & 0.010   &  -1.01 & 0.548  & -0.107 & 0.074  & 0.181  & 0.16           & 0.27          & -0.11       \\ 
53818  & 277.02 & 25.40  & 31.86$\pm$0.36   & 0.009   &  0.37  & 0.515  & 0.026  & 0.045  & 0.019  & -0.17          & 0.03          & -0.20       \\ 
54109  & 248.04 & 56.77  & 18.50$\pm$1.03   & 0.009   &  -0.05 & 0.621  & 0.139  & 0.152  & 0.013  & 0.02           & 0.03          & -0.01       \\ 
56832  & 260.42 & 62.79  & 28.53$\pm$0.49   & 0.005   &  0.12  & 0.675  & 0.136  & 0.221  & 0.085  & -0.03          & 0.11          & -0.14       \\ 
59199  & 290.66 & 37.12  & 66.95$\pm$0.15   & 0.005   &  -0.04 & 0.327  & -0.016 & 0.075  & 0.091  & 0.09           & 0.02          & +0.07       \\ 
66290  &  10.31 & 79.05  & 21.60$\pm$0.79   & 0.004   &  -0.50 & 0.426  & -0.103 & 0.002  & 0.105  & 0.17           & 0.12          & +0.05       \\ 
71192  & 107.22 & 48.50  & 33.86$\pm$0.59   & 0.002   &  -0.29 & 0.488  & -0.011 & 0.026  & 0.037  & 0.02           & 0.03          & -0.01       \\ 
73768  &  37.38 & 60.16  & 11.23$\pm$0.98   & 0.015   &  -0.41 & 0.465  & -0.061 & 0.013  & 0.074  & 0.10           & 0.08          & +0.02       \\ 
84489  & 350.92 & 1.42   & 31.81$\pm$0.51   & 0.014   &  -0.15 & 0.465  & -0.032 & 0.013  & 0.045  & 0.05           & 0.06          & -0.01       \\ 
86375  & 347.18 & -6.31  & 17.04$\pm$0.76   & 0.022   &  $-$   & 0.699  & 0.229  & 0.256  & 0.027  & -0.08          & 0.07          & -0.15       \\ 
92233  & 316.49 & -26.29 & 24.15$\pm$0.36   & 0.013   &  0.27  & 0.587  & 0.116  & 0.113  & -0.003 & -0.13          & 0.02          & -0.15       \\ 
93827  & 23.95  & -8.68  & 26.28$\pm$0.65   & 0.018   &  $-$   & 0.551  & -0.045 & 0.077  & 0.122  & 0.03           & 0.17          & -0.14       \\ 
96507  & 63.27  & 3.58   & 19.89$\pm$0.56   & 0.015   &  0.18  & 0.570  & 0.154  & 0.095  & -0.059 & -0.02          & 0.22          & -0.24       \\ 
99651  & 27.68  & -24.93 & 29.47$\pm$0.98   & 0.010   &  -0.63 & 0.703  & 0.133  & 0.262  & 0.129  & 0.09           & 0.17          & -0.08       \\ 
100925 & 12.18  & -33.06 & 51.22$\pm$0.54   & 0.006   &  0.00  & 0.729  & 0.264  & 0.303  & 0.039  & 0.03           & 0.07          & -0.04       \\ 
107020 & 53.33  & -37.89 & 24.25$\pm$0.92   & 0.009   &  -0.03 & 0.655  & 0.109  & 0.194  & 0.085  & -0.14          & 0.11          & -0.25       \\ 
108456 & 117.67 & 21.84  & 33.44$\pm$2.53   & 0.015   &  -0.25 & 0.502  & -0.026 & 0.036  & 0.062  & 0.07           & 0.05          & +0.02       \\ 
109090 & 83.87  & -22.08 & 18.00$\pm$0.76   & 0.011   &  -0.23 & 0.519  & -0.018 & 0.049  & 0.067  & 0.02           & 0.05          & -0.03       \\
113231 & 62.74  & -56.51 & 27.22$\pm$1.12   & 0.007   &  $-$   & 0.633  & 0.081  & 0.166  & 0.085  & -0.10          & 0.11          & -0.21       \\ 
118278 & 60.98  & -76.15 & 39.85$\pm$0.78   & 0.004   &  0.01  & 0.740  & 0.281  & 0.321  & 0.040  & -0.11          & 0.07          & -0.18       \\ 
\hline
\end{tabular}
\label{first}
\end{table*}

\begin{table*}
  \addtocounter{table}{-1}
  \renewcommand{\thetable}{\arabic{table}b}
\setlength{\tabcolsep}{2pt}
\center
\fontsize{8pt}{8pt}
\selectfont
\caption{Data for 39 field stars taken from the same source of stars used for the calibration of the the [$\alpha$/Fe] element and used for the application of the procedure. The symbols are the same as in Table 4a.} 
\label{second} 
\begin{tabular}{lcccccccccccccc}

   \hline
ID	        &$l$   & $b$         &$\pi$ &$E_d(B-V)$  &$(U-B)_0$  &$(B-V)_0$ & [Fe/H]  &[$\alpha$/Fe]$_{org}$ & $(U-B)_H$&$\delta$& [$\alpha$/Fe]$_{cal}$& $\Delta$[$\alpha$/Fe]\\
 	        &($^{o}$)& ($^{o}$)    &(mas)  &   (mag)    &   (mag) &  (dex)  &   (dex)      &(dex)   &  (mag)  & (mag)& (dex) & (dex)\\
\hline
 HIP 81170	&  11.66 & 27.71	& 22.17$\pm$1.35&	0.055&	 0.064	& 0.687	& -1.17	& 0.25	& 0.239&	0.174& 0.21 & +0.04\\
 HIP 93623	&  15.83 &-11.82	& 16.37$\pm$1.47&	0.012&	 0.039	& 0.641	& -0.74	& 0.27	& 0.180&        0.140& 0.20 & +0.07\\
 HIP 80837	&  19.30 & 33.05	& 23.41$\pm$0.79&	0.009&	-0.078	& 0.528	& -0.74	& 0.25	& 0.056&	0.134& 0.20 & +0.05\\
 HIP 84905	&  23.54 & 20.68	& 27.52$\pm$0.73&	0.019&	-0.029	& 0.560	& -0.50	& 0.19	& 0.085&	0.114& 0.16 & +0.03\\
 HIP 86443	&  26.77 & 17.03	&  8.31$\pm$1.67&	0.011&	-0.247	& 0.439	& -2.29	& 0.28	& 0.004&	0.251& 0.32 & -0.04\\
 HIP 72461	&  36.56 & 63.69	& 10.05$\pm$1.35&	0.014&	-0.239	& 0.411	& -2.48	& 0.32	& 0.002&	0.241& 0.30 & +0.02\\
 HIP 83489	&  37.35 & 31.30	& 13.43$\pm$1.41&	0.016&	 0.088	& 0.635	& -0.29	& 0.16	& 0.169&	0.081& 0.10 & +0.06\\
 HIP 74933	&  39.09 & 57.12	& 12.72$\pm$1.19&	0.017&	-0.017	& 0.552	& -0.39	& 0.04	& 0.078&	0.095& 0.11 & -0.07\\
 HIP 80003	&  39.52 & 42.85	&  8.16$\pm$3.37&	0.028&	 0.058	& 0.675	& $-$	& 0.15	& 0.221&	0.163& 0.24 & -0.09\\
 HIP 85912	&  42.48 & 25.53	& 26.56$\pm$0.39&	0.008&	-0.002	& 0.478	&-0.23	& 0.09	& 0.020&	0.022& 0.03 & +0.06\\
 HIP 87693	&  45.55 & 21.25	&  9.38$\pm$3.43&	0.022&	-0.206	& 0.418	&-2.07	& 0.32	& 0.002&	0.208& 0.26 & +0.06\\
 HIP 100279     &  49.00 &-16.69	& 10.46$\pm$1.60&	0.031&	-0.070	& 0.583	&-0.72	& 0.25	& 0.109&	0.179& 0.24 & +0.01\\
 HIP 97023	&  60.52 &  0.51        & 22.53$\pm$0.60&	0.010&	-0.010	& 0.555	&-0.48	& 0.14	& 0.081&	0.091& 0.10 & +0.04\\
 HIP 68321	&  61.86 & 73.96	&  4.86$\pm$1.36&	0.010&	-0.216	& 0.389	&-1.98	& 0.23	& 0.008&	0.224& 0.28 & -0.05\\
 HIP 78640	&  67.04 & 48.41        &  8.41$\pm$1.02&	0.006&	-0.203	& 0.468	&-1.43	& 0.19	& 0.015&	0.218& 0.28 & -0.09\\
 HIP 80	        &  82.75 &-70.65	& 13.68$\pm$1.22&	0.012&	-0.089	& 0.538	&-0.59	& 0.19	& 0.065&	0.154& 0.24 & -0.05\\
 HIP 113688     &  90.13 &-36.51	& 12.07$\pm$0.88&	0.024&	 0.073	& 0.586	&-0.14	&-0.01	& 0.112&	0.039& 0.05 & -0.06\\
 HIP 72407	&  98.57 & 52.33	& 10.65$\pm$0.73&	0.003&	 0.071	& 0.617	&-0.54	& 0.23	& 0.147&	0.076& 0.10 & +0.13\\
 HIP 74605	& 104.57 & 44.33	& 39.46$\pm$0.17&	0.003&	 0.067	& 0.531	&-0.02	& 0.03	& 0.059&       -0.008& 0.01 & +0.02\\
 HIP 14241	& 136.07 & 78.09	& 17.81$\pm$0.46&	0.004&	-0.023	& 0.476	& 0.04	&-0.01	& 0.019&	0.042& 0.03 & -0.04\\
 HIP 57450	& 145.97 & 63.25	&612.85$\pm$1.33&	0.006&	-0.154	& 0.554	& -1.23	& 0.15	& 0.080&	0.234& 0.21 & -0.06\\
 HIP 57939	& 168.53 & 73.78	&109.99$\pm$0.41& 	0.001&	 0.168	& 0.750	& -1.27	& 0.20	& 0.338&	0.170& 0.27 & -0.07\\
 HIP 48113	& 172.78 & 49.43	& 54.44$\pm$0.28&	0.001&	 0.143	& 0.612	& 0.04	& 0.06	& 0.141&       -0.002& 0.02 & +0.04\\
 HIP 14241	& 184.61 &-51.94	& 28.54$\pm$0.97&	0.012&	 0.108	& 0.658	&-0.45	& 0.24	& 0.198&	0.090& 0.12 & +0.12\\
 HIP 17147	& 189.77 &-43.12	& 39.12$\pm$0.56&	0.007&	-0.091	& 0.533	&-0.91	& 0.28	& 0.060&	0.151& 0.23 & +0.05\\
 HIP 24030	& 195.60 &-19.60	& 8.66$\pm$1.77&	0.023&	-0.147	& 0.497	&-1.10	& 0.24	& 0.032&	0.179& 0.25 & -0.01\\
 HIP 19814	& 198.48 &-37.12	& 13.55$\pm$2.08&	0.017&	 0.095	& 0.687	&-0.75	&-0.03	& 0.239&	0.143& 0.18 & -0.21\\
 HIP 44033	& 201.91 & 37.86	&  3.97$\pm$3.04&	0.022&	-0.157	& 0.548	& $-$	& 0.20	& 0.074&	0.231& 0.22 & -0.02\\
 HIP 52771	& 202.98 & 62.60	& 10.45$\pm$1.42&	0.012&	-0.233	& 0.488	&-1.85	& 0.15	& 0.026&	0.259& 0.30 & -0.15\\
 HIP 61545	& 206.07 & 86.70	&  3.21$\pm$2.06&	0.015&	-0.240	& 0.395	&-1.99	& 0.24	& 0.005&	0.245& 0.31 & -0.07\\
 HIP 57265	& 215.23 & 74.95	&  6.25$\pm$1.71&	0.014&	-0.147	& 0.465	&-0.93	& 0.03	& 0.013&	0.160& 0.19 & -0.16\\
 HIP 44124	& 232.63 & 25.95	& 11.76$\pm$1.59&	0.004&	-0.201	& 0.477	&-1.96	& 0.23	& 0.020&	0.221& 0.30 & -0.07\\
 HIP 57757	& 270.52 & 60.76	& 91.50$\pm$0.22&	0.002&	 0.101	& 0.550	& 0.13	& 0.03	& 0.076&       -0.025& 0.05 & -0.02\\
 HIP 58843	& 275.45 & 63.66	& 14.24$\pm$1.33&	0.008&	-0.059	& 0.579	& -0.84	& 0.26	& 0.105&	0.164& 0.22 & +0.04\\
 HIP 60632	& 288.36 & 63.42	&  9.05$\pm$1.11&	0.011&	-0.234	& 0.421	& -1.75	& 0.23	& 0.002&	0.236& 0.30 & -0.07\\
 HIP 5054	& 291.21 &-77.34	& 24.17$\pm$0.61&	0.003&	-0.012	& 0.600	& -0.62	& 0.27	& 0.127&	0.139& 0.20 & +0.07\\
 HIP 108736     & 341.43 &-49.37	& 27.95$\pm$0.55&	0.004&	 0.569  & 0.565	& -0.29	& 0.18	& 0.090&	0.068& 0.05 & +0.13\\
 HIP 92288	& 345.38 &-20.23	& 11.99$\pm$1.30&	0.012&	 0.106	& 0.630	& -0.21	& 0.15	& 0.163&	0.057& 0.07 & +0.08\\
 HIP 100412     & 349.13 &-34.67	& 32.24$\pm$0.47&	0.005&	 0.551  & 0.546	& -0.26	& 0.06	& 0.072&	0.096& 0.11 & -0.05\\
\hline
\end{tabular}
\\
\end{table*}

\begin{table*}
  \addtocounter{table}{-1}
  \renewcommand{\thetable}{\arabic{table}c}
\setlength{\tabcolsep}{3pt}
\center
\tiny
\caption{Data for 25 clusters whose $B-V$ and $U-B$ colour indices, and the colour excess $E(B-V)$, 
are taken from \cite{Harris10}. While, the iron elements $[Fe/H]$ and the alpha elements [$\alpha$/Fe] are provided from \cite{Dias16}.The columns give: cluster name, colour excess $E(B-V)$, $B-V$ and $U-B$ colour indices, de-reddened $(B-V)_0$ and $(U-B)_0$ colour indices, iron element [Fe/H], original alpha element [$\alpha$/Fe]$_{org}$, ultra-violet colour index for the Hyades cluster $(U-B)_{H}$, ultra-violet excess $\delta$, estimated alpha element [$\alpha$/Fe]$_{cal}$, and the residual $\Delta$[$\alpha$/Fe].} 
\label{third} 
\begin{tabular}{lccccccccccc}
\hline
Cluster & $E(B-V)$ & $B-V$  & $U-B$  & $(B-V)_0$ & $(U-B)_0$ & [Fe/H] & [$\alpha$/Fe]$_{org}$ & $(U-B)_H$ & $\delta$   & [$\alpha$/Fe]$_{cal}$ & $\Delta$[$\alpha$/Fe] \\
        &    (mag) & (mag)  & (mag)  & (mag)     & (mag)     & (dex)    & (dex)               & (mag)     &   (mag)    &          (dex)      & (dex) \\
\hline
NGC 2298         & 0.14   & 0.75 & 0.17 & 0.61   & 0.07   & -1.95      & 0.19          & 0.14   & 0.07  & 0.09          & 0.097       \\
NGC 2808         & 0.22   & 0.92 & 0.28 & 0.70   & 0.12   & -1.06      & 0.24          & 0.26   & 0.14  & 0.18          & 0.062       \\
NGC 3201         & 0.24   & 0.96 & 0.38 & 0.72   & 0.20   & -1.51      & 0.22          & 0.29   & 0.08  & 0.12          & 0.096       \\
NGC 4372         & 0.39   & 1.10 & 0.31 & 0.71   & 0.02   & -2.20      & 0.21          & 0.27   & 0.25  & 0.27          &-0.056      \\
NGC 4590(M68)    & 0.05   & 0.63 & 0.04 & 0.58   & 0.00   & -2.20      & 0.19          & 0.11   & 0.10  & 0.14          & 0.049       \\
NGC 5694         & 0.09   & 0.69 & 0.08 & 0.60   & 0.01   & -1.98      & 0.17          & 0.13   & 0.11  & 0.16          & 0.012       \\
NGC 5634         & 0.05   & 0.67 & 0.09 & 0.62   & 0.05   & -1.75      & 0.20          & 0.15   & 0.10  & 0.13          & 0.067       \\
NGC 5824         & 0.13   & 0.75 & 0.12 & 0.62   & 0.03   & -1.99      & 0.24          & 0.15   & 0.12  & 0.18          & 0.064       \\
NGC 5897         & 0.09   & 0.74 & 0.08 & 0.65   & 0.01   & -1.97      & 0.23          & 0.19   & 0.17  & 0.26          &-0.031      \\
NGC 5904(M5)     & 0.03   & 0.72 & 0.17 & 0.69   & 0.15   & -1.25      & 0.24          & 0.24   & 0.09  & 0.13          & 0.106       \\
NGC 5946         & 0.54   & 1.29 & 0.45 & 0.75   & 0.05   & -1.54      & 0.22          & 0.34   & 0.29  & 0.42          &-0.196      \\
NGC 6121(M4)     & 0.35   & 1.03 & 0.43 & 0.68   & 0.17   & -1.01      & 0.27          & 0.23   & 0.06  & 0.10          & 0.172       \\
NGC 6171(M107)   & 0.33   & 1.10 & 0.69 & 0.77   & 0.45   & -0.95      & 0.20          & 0.37   &-0.07  & 0.01          & 0.185      \\
NGC 6254(M10)    & 0.28   & 0.90 & 0.23 & 0.62   & 0.02   & -1.56      & 0.21          & 0.15   & 0.13  & 0.18          & 0.032       \\
NGC 6284         & 0.28   & 0.99 & 0.40 & 0.71   & 0.19   & -1.07      & 0.27          & 0.27   & 0.08  & 0.12          & 0.151       \\
NGC 6355         & 0.77   & 1.48 & 0.72 & 0.71   & 0.14   & -1.46      & 0.27          & 0.27   & 0.14  & 0.18          & 0.094       \\
NGC 6366         & 0.71   & 1.44 & 0.97 & 0.73   & 0.43   & -0.61      & 0.30          & 0.30   &-0.13  & 0.07          & 0.232       \\
NGC 6397         & 0.18   & 0.73 & 0.12 & 0.55   &-0.01   & -2.07      & 0.23          & 0.08   & 0.09  & 0.09          & 0.137       \\
NGC 6453         & 0.64   & 1.31 & 0.68 & 0.67   & 0.20   & -1.54      & 0.16          & 0.21   & 0.02  & 0.03          & 0.133       \\
NGC 6558         & 0.44   & 1.11 & 0.58 & 0.67   & 0.25   & -1.01      & 0.23          & 0.21   &-0.04  & 0.02          & 0.205       \\
NGC 6656(M22)    & 0.34   & 0.98 & 0.28 & 0.64   & 0.03   & -1.92      & 0.22          & 0.17   & 0.15  & 0.21          & 0.006       \\
NGC 6752         & 0.04   & 0.66 & 0.07 & 0.62   & 0.04   & -1.57      & 0.22          & 0.15   & 0.11  & 0.15          & 0.067       \\
NGC 6864(M75)    & 0.16   & 0.87 & 0.28 & 0.71   & 0.16   & -1.00      & 0.22          & 0.27   & 0.11  & 0.15          & 0.071       \\
NGC 7006         & 0.05   & 0.75 & 0.06 & 0.70   & 0.02   & -1.69      & 0.25          & 0.26   & 0.23  & 0.26          &-0.006      \\
NGC 7078(M15)    & 0.10   & 0.68 & 0.06 & 0.58   &-0.01   & -2.23      & 0.24          & 0.11   & 0.12  & 0.17          & 0.074       \\
\hline
\end{tabular}
\end{table*}

\begin{table*}
  \addtocounter{table}{-1}
  \renewcommand{\thetable}{\arabic{table}d}
\setlength{\tabcolsep}{1.2pt}
\center
\tiny
\caption{Data for 19 clusters whose $B-V$ and $U-B$ colour indices are taken from \cite{Hanes85}. 
The colour excess $E(B-V)$ is taken from \cite{Harris10}. The symbols ``con.'' and ``not'' in the Remarks column indicate that the errors in $B-V$ or $U-B$ are considered or not considered, respectively. The residuals in the last column correspond to ones in Table 4c, for comparison.}
\label{fourth} 

\begin{tabular}{lccccccccccccc}
\hline
Cluster& $E(B-V)$ & $B-V$  & $U-B$ & [Fe/H] & [$\alpha$/Fe] & $(B-V)_0$ & $(U-B)_0$ & $(U-B)_H$ & $\delta$  & [$\alpha$/Fe]$_{cal}$ & $\Delta$[$\alpha$/Fe] & Remarks & $\Delta$[$\alpha$/Fe]$_{4c}$ \\
       & (mag)    & (mag)  & (mag) &  (dex)   & (dex) & (mag) & (mag) & (mag) & (mag)  & (dex) & (dex) &   & (dex) \\
\hline
NGC 2298         & 0.14   & 0.720$\pm$0.040 & 0.27$\pm$0.17   & -1.95      & 0.19       & 0.62   & -0.002  & 0.150  & 0.152  & 0.212         & -0.022      & con.    & 0.097  \\
NGC 2808         & 0.22   & 0.900           & 0.18            & -1.06      & 0.24       & 0.68   & -0.040  & 0.228  & 0.268  & 0.274         & -0.034      & --      & 0.062  \\
NGC 3201         & 0.24   & 0.980           & 0.45$\pm$0.09   & -1.51      & 0.22       & 0.74   & 0.210   & 0.321  & 0.111  & 0.175         & 0.045       & not     & 0.096  \\
NGC 4590(M68)    & 0.05   & 0.660           & 0.05            & -2.20      & 0.19       & 0.61   & 0.000   & 0.139  & 0.139  & 0.196         & -0.006      & --      & 0.049  \\
NGC 5634         & 0.05   & 0.680           & 0.31$\pm$0.10   & -1.75      & 0.20       & 0.63   & 0.260   & 0.163  & -0.097 & 0.119         & 0.081       & not     & 0.067  \\
NGC 5694         & 0.09   & 0.690           & 0.01$\pm$0.05   & -1.98      & 0.17       & 0.60   & -0.130  & 0.127  & 0.257  & 0.195         & -0.025      & con.    & 0.012  \\
NGC 5824         & 0.13   & 0.760           & 0.02            & -1.99      & 0.24       & 0.63   & -0.110  & 0.163  & 0.273  & 0.359         & -0.119      & --      & 0.064  \\
NGC 5904(M5)     & 0.03   & 0.710           & 0.07            & -1.25      & 0.24       & 0.68   & 0.040   & 0.228  & 0.188  & 0.223         & 0.017       & --      & 0.106  \\
NGC 5927         & 0.45   & 1.270           & 0.85$\pm$0.12   & -0.21      & 0.30       & 0.82   & 0.280   & 0.469  & 0.189  & 0.301         & -0.001      & con.    & --     \\
NGC 5946         & 0.54   & 1.100           & 0.41$\pm$0.09   & -1.54      & 0.22       & 0.56   & -0.130  & 0.085  & 0.215  & 0.253         & -0.033      & not     & --     \\
NGC 6121(M4)     & 0.35   & 1.030           & 0.32$\pm$0.05   & -1.01      & 0.27       & 0.68   & -0.030  & 0.228  & 0.258  & 0.269         & 0.001       & not     & 0.172  \\
NGC 6171(M107)   & 0.33   & 1.150$\pm$0.030 & 0.67$\pm$0.11   & -0.95      & 0.20       & 0.82   & 0.340   & 0.469  & 0.129  & 0.204         & -0.004      & not     & 0.185  \\
NGC 6254(M10)    & 0.28   & 0.870           & 0.59$\pm$0.10   & -1.56      & 0.21       & 0.59   & 0.210   & 0.116  & -0.094 & 0.151         & 0.059       & con.    & 0.032  \\
NGC 6284         & 0.28   & 0.990           & 0.38$\pm$0.05   & -1.07      & 0.27       & 0.71   & 0.050   & 0.273  & 0.223  & 0.249         & 0.021       & con.    & 0.151  \\
NGC 6355         & 0.77   & 1.500$\pm$0.030 & 0.62$\pm$0.11   & -1.46      & 0.27       & 0.70   & -0.040  & 0.258  & 0.298  & 0.282         & -0.012      & con.    & 0.094  \\
NGC 6397         & 0.18   & 0.740           & 0.12            & -2.07      & 0.23       & 0.56   & -0.060  & 0.085  & 0.145  & 0.224         & 0.006       & --      & 0.137  \\
NGC 6441         & 0.47   & 1.300           & 0.79            & -0.41      & 0.26       & 0.83   & 0.320   & 0.490  & 0.170  & 0.270         & -0.010      & --      & --     \\
NGC 6752         & 0.04   & 0.710           &-0.01            & -1.57      & 0.22       & 0.67   & -0.050  & 0.214  & 0.264  & 0.357         & -0.137      & --      & 0.067  \\
NGC 6864(M75)    & 0.16   & 0.880           & 0.18            & -1.00      & 0.22       & 0.72   & 0.020   & 0.288  & 0.268  & 0.274         & -0.054      & --      & 0.071  \\    
\hline
\end{tabular}

\end{table*}

\begin{table*}
\renewcommand{\thetable}{\arabic{table}a}
\setlength{\tabcolsep}{3pt}
\center
\caption{Comparison of the $\alpha$ elements determined in three different studies.}
\label{first}  
\begin{tabular}{lccccccccc}
\hline
\multicolumn{1}{c}{Star}&\multicolumn{2}{c}{HD108177}&\multicolumn{2}{c}{HD149414}&\multicolumn{3}{c}{BD+0203375}&\multicolumn{2}{c}{BD+2003603}\\ \cline{2-10} \cline{2-3} \cline{6-8}
\multicolumn{1}{c}{}& \multicolumn{1}{c}{Bai04}&\multicolumn{1}{c}{F00}&\multicolumn{1}{c}{Bai04}&\multicolumn{1}{c}{F00}& \multicolumn{1}{c}{Bai04}&\multicolumn{1}{c}{F00}&\multicolumn{1}{c}{S02}&\multicolumn{1}{c}{Bai04}&\multicolumn{1}{c}{F00}\\
\hline
$[Mg/Fe]$ 	& 0.09	& 0.43  & 0.22 & 0.41 & 0.24 & 0.52 & 0.34 & 0.36 & 0.46\\
$[Ca/Fe]$   	& 0.32  & 0.36  & 0.45 & 0.27 & 0.36 & 0.38 & 0.39 & 0.28 & 0.37\\
$[Ti/Fe]$    	& 0.47  & 0.44  & 0.36 & 0.39 & 0.48 & 0.42 & 0.38 & 0.53 & 0.48\\
$[\alpha/Fe]$   & 0.20  & 0.29  & 0.23 & 0.24 & 0.25 & 0.32 & 0.25 & 0.30 & 0.32\\
\hline
\hline
\multicolumn{1}{c}{Star}               &             \multicolumn{2}{c}{BD+26 2606}			  &              \multicolumn{2}{c}{BD+29 2091}             &            \multicolumn{2}{c}{BD+34 2476}              & \multicolumn{3}{c}{BD+42 2667}                                                   \\ 
\hline
\multicolumn{1}{c}{}                   & \multicolumn{1}{c}{Bai04} 	& \multicolumn{1}{c}{F00}       & \multicolumn{1}{c}{Bai04}   & \multicolumn{1}{c}{F00} & \multicolumn{1}{c}{Bai04} & \multicolumn{1}{c}{S02}  & \multicolumn{1}{c}{Bai04} & \multicolumn{1}{c}{F00} & \multicolumn{1}{c}{S02} \\ 
\hline
$[Mg/Fe]$ 	& 0.25  & 0.42  & 0.11 & 0.38  & 0.17    & 0.27  & 0.09 & 0.41  & 0.36  \\
$[Ca/Fe]$  	& 0.28  & 0.37  & 0.24 & 0.42  & 0.34    & 0.49  & 0.32 & 0.38  & 0.33  \\
$[Ti/Fe]$  	& 0.63  & 0.52  & 0.38 & 0.54  & 0.43    & 0.44  & 0.39 & 0.36  & 0.31  \\
$[\alpha/Fe]$  	& 0.30  & 0.32  & 0.13 & 0.33  & 0.20    & 0.29  & 0.16 & 0.26  & 0.21  \\ 
\hline
\end{tabular}
\\
(1) Bai04: \citet{Bai04}, (2) F00: \citet{Fulbright00}, (3) S02: \citet{Stephens02}.\\  
\end{table*}

\begin{table*}
\addtocounter{table}{-1}
  \renewcommand{\thetable}{\arabic{table}b}
\setlength{\tabcolsep}{3pt}
\center
\caption{Comparison of the $\delta$ elements determined in different studies.}  
\label{second}
\begin{tabular}{lccccccccc}
\hline
\multicolumn{1}{c}{Star}&\multicolumn{2}{c}{[$\alpha$/Fe]}&\multicolumn{1}{c}{Star}&\multicolumn{2}{c}{[$\alpha$/Fe]}& & & & \\
\multicolumn{1}{c}{}& \multicolumn{1}{c}{V04}&\multicolumn{1}{c}{B14}&\multicolumn{1}{c}{}&\multicolumn{1}{c}{V04}&\multicolumn{1}{c}{N10}& & & & \\
\hline                              
HIP 47048	&0.06	& 0.14	  & 	  HIP 19797	&        0.30  &    0.20& & & & \\
HIP 107975	&0.18	& 0.11	  &       HIP 57265	&        0.13  &    0.03& & & & \\
HIP 60632	&0.34	& 0.25	  &       HIP 94449	&        0.33  &    0.26& & & & \\
HIP 72673	&0.19	& 0.14	  &       HIP 24030	&        0.27  &    0.24& & & & \\
HIP 110341	&0.09	& 0.04	  &       HIP 58229	&        0.13  &    0.09& & & & \\
HIP 3026	&0.23	& 0.15	  &       HIP 7459	&        0.10  &    0.15& & & & \\
HIP 2711       &-0.05	& 0.06	  &       HIP 36818	&       -0.03  &    0.04& & & & \\
HIP 78640	&0.30	& 0.27 	  &  	  	$-$	& 	 $-$   &    $-$ & & & & \\ \cline{4-10}
HIP 9085	&0.07	& 0.07	  &                     &	 V04   &     R06& & & & \\ \cline{4-10} 
HIP 22632	&0.19	& 0.25    &       HIP 88039	&	 0.30  &    0.24& & & & \\
HIP 699	   	&0.16	& 0.05    &       HIP 52771	&	 0.36  &    0.11& & & & \\
HIP 910		&0.09	& 0.08    &       HIP 31188     &        0.14  &    0.05& & & & \\
HIP 96536	&0.07	& 0.00    &       HIP 99938     &        0.30  &    0.16& & & & \\
HIP 58145	&0.18	& 0.26	  &       HIP 10449     &        0.27  &    0.14& & & & \\
HIP 3909	&0.06   &-0.02    &       HIP 85373     &        0.25  &    0.10& & & & \\
HIP 22325	&0.08   &-0.01    &       HIP 85378     &        0.30  &    0.24& & & & \\
HIP 16404	&0.37	& 0.17    &       HIP 25860     &        0.25  &    0.18& & & & \\
HIP 67863	&0.20	& 0.27	  &          $-$	&	$-$    &     $-$& & & & \\
HIP 5054	&0.22	& 0.27	  &          $-$	&	$-$    &     $-$& & & & \\
HIP 7276	&0.07   &-0.01	  &          $-$	&	$-$    &     $-$& & & & \\		
\hline
\end{tabular}
\\
Note: V04: \citet{Venn04}, B14: \citet{Bensby14}, N10: \citet{Nissen10}, R06: \citet{Reddy06}.\\  
\end{table*}

\begin{table*}[]
\centering
\caption{Percentages of the stars ($\%$) with (absolute) residuals less than 0.15 and 0.20 in four panels of Fig. 5.}

\begin{tabular}{ccccc}
\hline
Panel $\to$ &(a)& (b)& (c) &(d)\\
\hline
$|\Delta$[Fe/H]$|<0.15$ & 63 &  92 &80 &100\\
$|\Delta$[Fe/H]$|<0.20$ & 79 & 100 &92 &100\\
\hline
\end{tabular}
\end{table*}

\end{document}